\documentclass[runningheads]{llncs}

\newcommand{\F}{\mathbb{F}}

\usepackage{amsfonts}
\usepackage{amsmath}
\usepackage{float}

\usepackage[T1]{fontenc}
\usepackage{graphicx}
\begin{document}
\title{Transformation of the discrete logarithm problem over $\F_{2^n}$ to the QUBO problem using normal bases}
\titlerunning{Transformation of the DLP over $\F_{2^n}$}
%
\author{Michał Wroński\orcidID{0000-0002-8679-9399} \and
Mateusz Leśniak\orcidID{0009-0001-0092-2975}}
\authorrunning{M. Wroński, M. Leśniak}
%
\institute{NASK - National Research Institute, Warsaw, Poland\\
\{michal.wronski, mateusz.lesniak\}@nask.pl}
\maketitle              
\begin{abstract}
Quantum computations are very important branch of modern cryptology. According to the number of working physical qubits available in general-purpose quantum computers and in quantum annealers, there is no coincidence, that nowadays quantum annealers allow to solve larger problems. In this paper we focus on solving discrete logarithm problem (DLP) over binary fields using quantum annealing. It is worth to note, that however solving DLP over prime fields using quantum annealing has been considered before, no author, until now, has considered DLP over binary fields using quantum annealing. Therefore, in this paper, we aim to bridge this gap. We present a polynomial transformation of the discrete logarithm problem over binary fields to the Quadratic Unconstrained Binary Optimization (QUBO) problem, using approximately $3n^2$ logical variables for the binary field $\mathbb{F}_{2^n}$. In our estimations, we assume the existence of an optimal normal base of II type in the given fields. Such a QUBO instance can then be solved using quantum annealing.

\end{abstract}
\section{Introduction}
Shor's algorithm \cite{Sho94} has long been considered the most promising method for solving classical cryptography problems such as integer factorization, the discrete logarithm problem in finite fields, and the elliptic curve discrete logarithm problem.

In 2023, Shor's algorithm was improved by Regev \cite{regev2023efficient}, and later by Ragavan and Vaikuntanathan \cite{Ragavan2023} in the case of integer factorization, and then by Ekerå and Gärtner in the case of the discrete logarithm problem in finite fields \cite{ekerå2023extending}.

However, it is worth noting that other quantum methods have become more prominent with the existing quantum hardware. Taking this into account, quantum annealing is now considered the most practical method of quantum computation, which allows for computing some instances of the discrete logarithm problem (DLP) over finite fields that are larger than those solvable using Shor's algorithm \cite{Wro22DLP}, \cite{wronski2024transformation}.

It is worth noting that the application of quantum annealing to cryptanalysis has been considered previously. The first paper, presented by Jiang in \cite{Jiang2018}, demonstrated the use of quantum annealing for the cryptanalysis of the factorization problem. Subsequently, other cryptanalysis problems have been explored, such as breaking the discrete logarithm \cite{wronski2021index}, \cite{Wro22DLP}, \cite{zolnierczyk2023searching}, \cite{wronski2024transformation}, \cite{wronski2024base} and attacking symmetric ciphers \cite{burek2022algebraic}, \cite{burek2022speck}, \cite{cryptoeprint:2023/1502}, \cite{burek2023searching}, \cite{LesniakE0}, \cite{morse2024compactquboencodingcomputational}.

As shown above, while computing the discrete logarithm problem (DLP) over prime fields using quantum annealing has been considered before, no author has addressed DLP over binary fields using quantum annealing.

In this paper, we aim to bridge this gap. We present a polynomial transformation of the discrete logarithm problem over binary fields to the Quadratic Unconstrained Binary Optimization (QUBO) problem, using approximately $3n^2$ logical variables for the binary field $\mathbb{F}_{2^n}$. In our estimations, we assume the existence of an optimal normal base of II type in the given fields. Such a QUBO instance can then be solved using quantum annealing.

Interestingly, this transformation requires more logical variables than a similar transformation for DLP over a prime field $\mathbb{F}_p$, whose bit-length is equal to $n$. The main reason is the complexity of transforming binary field arithmetic into a pseudo-Boolean function. Therefore, there are many additional variables, making such a transformation less efficient than in the case of prime fields.

\section{DLP in $\mathbb{F}_{2^n}$ using optimal normal bases}

This section focuses on the fields $\mathbb{F}_{2^n}$. We assume the use of such $\mathbb{F}_{2^n}$ fields where an optimal normal basis exists.

For the field $\mathbb{F}_{2^n}$, the normal basis consists of elements $\{\alpha^{2^0}, \alpha^{2^1}, \dots, \alpha^{2^{n-1}}\}$, whereas in the commonly used polynomial basis, elements form the set\linebreak $\{1, \beta, \dots, \beta^{n-1}\}$. We assume here that the field $\mathbb{F}_{2^n}$ is generated by the irreducible polynomial $f(t)$ of degree $n$.

It is worth noting that if in the given field exists the optimal normal basis of the II type, then it is always possible to find such a polynomial $f(t)$ of degree $n$ that $\alpha = t$ is the generator of the normal basis in the field $\mathbb{F}_{2^n}$.

Such an irreducible polynomial can be constructed recursively using Dickson polynomials in the following manner \cite{mullin2005dickson}:

\begin{equation}
\label{DickPol}
\begin{cases}
    f_0(t) = 1,\\
    f_1(t) = t + 1,\\
    f_n(t) = t \cdot f_{n-1}(t) - f_{n-2}(t), \quad n \geq 2.
\end{cases}
\end{equation}

A normal basis is optimal if its multiplication matrix $T$ consists of $2n-1$ nonzero elements.

For the simplicity of our estimations, in the subsequent sections, we assume that an optimal normal basis exists for the given binary field $\mathbb{F}_{2^n}$. Therefore, in the multiplication matrix $T$, in $n-1$ rows (columns), there are just two elements "1", and in one row (column), there occurs just one "1". This assumption will be important in the analysis of the complexity of our problem.

Let us make the following assumption: the field $\mathbb{F}_{2^n}$ is generated by the irreducible polynomial $f(t)$, given by the Dickson polynomial, for such a field an optimal normal basis of II type exists, and the generator of the multiplicative subgroup of this field is $t$.

In such a case, one can define the multiplication matrix $T^{(0)}$, which allows one to obtain the least significant bit $c_0$ of the resulting register $C$. However, by making rotations of columns and rows, it is also possible to obtain other bits of register $C$ (see \cite{amin2006hardware}).

Therefore, the multiplication of any two elements $A, B \in \mathbb{F}_{2^n}$ (presented as vectors) is performed in the following manner:

\begin{equation} 
   C = A \cdot T^{(0)} \cdot B^T.
\end{equation}
Then the product $C$, generally, may be given by the following system of equations:

\begin{equation}
\label{EqNorGen}
\begin{array}{rl}
    c_k = & a_{(i_0+k \mod n)} b_{(j_0+k \mod n)}\\
    + & (a_{(i_1+k \mod n)} b_{(j_1+k \mod n)}\\
    + & a_{(j_1+k \mod n)} b_{(i_1+k \mod n)})\\
    + & \dots\\
    + & (a_{(i_{n-2}+k \mod n)} b_{(j_{n-2}+k \mod n)}\\
    + &a_{(j_{n-1}+k \mod n)} b_{(i_{n-1}+k \mod n)}).
\end{array}
\end{equation}

Now, let us assume that the generator of the multiplicative subgroup is $t$.

We begin the main part of this section by defining the discrete logarithm problem, similarly as in \cite{Wro22DLP}, \cite{wronski2024transformation}:
\begin{equation}
\label{DLPeq1}
t^y = h,
\end{equation}
in the multiplicative subgroup of the field \(\mathbb{F}_{2^n}\), so \(t, h \in \mathbb{F}_{2^n}^*\) and \(y \in \{1, \dots, 2^n-1\}\). This problem is equivalent to:
\begin{equation}
\label{DLPeq2}
t^y \equiv h \mod f(t),
\end{equation}
for elements \(t, h \in \mathbb{F}_{2^n}^*\) and \(y \in \{1, \dots, 2^n-1\}\).

Let us note that the bit length of \(2^n-1\) equals $n$. Noting that \(y\) can be written using \(n\) bits, and if \(y = 2^{n-1}u_{n} + \dots + 2u_2 + u_1\), where \(u_1, \dots, u_n\) are binary variables, then
\begin{equation}
\label{Eq1}
t^y = t^{2^{n-1}u_{n}} \cdots t^{2u_2} t^{u_1}.
\end{equation}

Let us also note that
\begin{equation}
A \cdot t^{2^{i-1}u_{i}} =
\begin{cases}
A, & \text{if } u_i = 0, \\
A \cdot t^{2^{i-1}}, & \text{if } u_i = 1.
\end{cases}
\end{equation}

So now let us assume that we have to perform the multiplication of $C = A \cdot t^{2^{l-1} u_l}$.

Using Equation \eqref{EqNorGen}, one obtains the following result:

\begin{equation}
\label{GenBit}
\begin{array}{rl}
    c_k = & a_{i} u_l + a_{j} u_l + a_k(1 - u_l).
    \end{array}
\end{equation}

Let us note that the equation above is correct. When $u_l = 1$, then $C = A \cdot t^{2^{l-1}}$, and therefore in the normal basis representation, only bit $b_l$ is equal to $1$, and others are equal to $0$. In the opposite situation, when $u_l = 0$, then $C = A \cdot 1$. However, in normal bases representation, $1 = t + t^2 + \dots + t^{2^{n-1}}$. Note that in Equation \eqref{GenBit}, both situations are considered. If $u_l = 0$, then $c_k = a_k$ for every $k = \overline{0, n-1}$, which results in $C = A$, and therefore it is equivalent to multiplication by $1$. Otherwise, if $u_l = 1$, then $c_k = a_{i} u_l + a_{j} u_l$, where at least one of $a_i, a_j$ is non-zero (we assume that we always perform multiplication by of non-zero elements). Therefore, $C$ is the result of multiplication of register $A$ by $t^{2^{l-1}}$.

\section{Example analysis for DLP over $\mathbb{F}_{2^5}$ and its generalization}

To better illustrate the transformation of the discrete logarithm problem over binary fields to the QUBO problem, we will demonstrate this process using the small degree extension field $\mathbb{F}_{2^5}$. Using such a small field should clarify the general method for transforming DLP over binary fields to the QUBO problem.

Consider the given field $\mathbb{F}_{2^5}$. For such a field, there exists an optimal normal basis of type II \cite{amin2006hardware}.

The irreducible polynomial $f(t)$, which generates the field $\mathbb{F}_{2^5}$, is given as $f(t)=t^5+t^4+t^2+t+1$.

When using the normal basis, the multiplication matrix $T^{(0)}$ is given \cite{amin2006hardware} as: 

\begin{equation}
   T^{(0)}= \begin{bmatrix}
0 & 1 & 0 & 0 & 0\\
1 & 0 & 0 & 1 & 0\\
0 & 0 & 0 & 1 & 1\\
0 & 1 & 1 & 0 & 0\\
0 & 0 & 1 & 0 & 1
\end{bmatrix}.
\end{equation}

Now let $A,B,C \in \mathbb{F}_{2^5}$, where $C=A \cdot B$. Then, the product $C$ is given by

\begin{equation}
\begin{array}{rl}
    c_0=&a_4 b_4 + (a_0 b_1 + a_1 b_0) + (a_1 b_3 + a_3 b_1) \\
    +&(a_2 b_4 + a_4 b_2) + (a_2 b_3 + a_3 b_2),\\
    \\
    c_1=&a_0 b_0 + (a_1 b_2 + a_2 b_1) + (a_2 b_4 + a_4 b_2) \\
    +&(a_3 b_0 + a_0 b_3) + (a_3 b_4 + a_4 b_3),\\
    \\
    c_2=&a_1 b_1 + (a_2 b_3 + a_3 b_2) + (a_3 b_0 + a_0 b_3)\\
    +&(a_4 b_1 + a_1 b_4) + (a_4 b_0 + a_0 b_4),\\
    \\
    c_3=&a_2 b_2 + (a_3 b_4 + a_4 b_3) + (a_4 b_1 + a_1 b_4)\\
    +&(a_0 b_2 + a_2 b_0) + (a_0 b_1 + a_1 b_0),\\
    \\
    c_4=&a_3 b_3 + (a_4 b_0 + a_0 b_4) + (a_0 b_2 + a_2 b_0)\\
    +&(a_1 b_3 + a_3 b_1) + (a_1 b_2 + a_2 b_1).  
\end{array}
\end{equation}

Now, let us assume that the generator of the multiplicative subgroup is $t$.
We begin the main part of this section by defining the discrete logarithm problem, similarly as presented in Equations \eqref{DLPeq1} and \eqref{DLPeq2}.

For any $A \in \mathbb{F}_{2^5}$ given as $a_4 t^{2^4} + a_3 t^{2^3} + a_2 t^{2^2} + a_1 t^{2^1} + a_0 t^{2^0}$, let us set, for example, $B = t^{2^0 u_0}$. In such a case, the result $C$ of the multiplication of $A \cdot B$, will be given as:

\begin{equation}
\begin{array}{rl}
     c_0=&a_1 u_0 + a_0(1+u_0),\\
    c_1=&a_0 u_0 + a_3 u_0 + a_1(1+u_0),\\
    c_2=&a_3 u_0 + a_4 u_0 + a_2(1+u_0),\\
    c_3=&a_2 u_0 + a_1 u_0 + a_3(1+u_0),\\
    c_4=&a_4 u_0 + a_2 u_0 + a_4(1+u_0).
\end{array}
\end{equation}

Detailed analysis shows that for every field $\mathbb{F}_{2^n}$, the resulting register $C$ will have a similar form for multiplying $A$ by any $B = t^{2^i u_i}$, for $\overline{0,n-1}$. More precisely, $n-1$ bits of register $C$ will consist of $4$ monomials: 2 monomials of degree $2$ occur because in the multiplication matrix $T^{(0)}$ there are two "1"s, one monomial of degree $2$, and one monomial of degree $1$, because there is multiplication of $a_i$ by $(1+u_j)$, for $\overline{0, n-1}$. One bit of register $C$ will consist of $3$ monomials: 1 monomial of degree $2$ occurs because in the multiplication matrix $T^{(0)}$ there is one "1", and because there is multiplication of $a_i$ by $(1+u_j)$, for $\overline{0, n-1}$, one monomial of degree $2$, and one monomial of degree $1$.

Therefore, we may estimate the total number of variables required for transforming DLP over $\mathbb{F}_{2^n}$ to the QUBO problem. We use the same decomposition tree as in \cite{Wro22DLP}.

For a single node, there will be necessary $n$ binary variables for register $C$. Moreover, there are also necessary $n$ new variables for linearization (note that monomials of degree two of the form $u_i a_j$, where $j = \overline{0, n-1}$ will occur). There will also be necessary $2(n-1)$ new variables for $k$ (this is necessary for $n-1$ bits), and for one bit of register $C$, one additional variable for $k$ will be necessary. Of course, one variable is necessary to represent $u_i$. Summing up, there will be necessary $4n$ variables for a single node. As we have approximately $n$ nodes, the total number of variables will equal approximately $4n^2$.

However, the amount of variables described above ($4n^2$) may be lowered. Let us note that each of the single equations $c_i$ can be transformed into a pseudo-boolean function in the following manner (let us take, for example, $c_1$): 
\begin{equation*}
    c_1 + a_0 u_0 + a_3 u_0 + a_1(1 + u_0) = 0.
\end{equation*}
As the maximal value of the left side of the equation is equal to $5$, two new variables are necessary for $k$. However, note that in binary notation, the equation above is equivalent to 
\begin{equation*}
    -c_1 + a_0 u_0 + a_3 u_0 + a_1(1 - u_0) = 0.
\end{equation*} 
Now one can transform this equation into the pseudo-boolean function as \begin{equation*}
    -c_1 + a_0 u_0 + a_3 u_0 + a_1(1 - u_0) - 2k_1 = 0,
\end{equation*} where $k_1 \in \{0,1\}$. Why is it possible? Let's observe that the maximal value of $-c_1 + a_0 u_0 + a_3 u_0 + a_1(1 - u_0)$ is equal to $2$, and the minimal value is equal to $-1$. It means that $0 \leq k_1 \leq 1$. Therefore, we require only one bit for multiplicity representation instead of two bits, as was described above.

It is important to see that a similar trick may be done for all fields $\mathbb{F}_{2^n}$, for which optimal normal bases exist. In such a case, transforming the DLP problem over $\mathbb{F}_{2^n}$ will require $3n^2$ logical variables. However, whether this transformation may be obtained using fewer variables is unknown.




\section{Working example}

In this section the practical example of application of our method will be presented.

\subsection{Normal bases and multiplication matrix definition over $\F_{2^3}$}

According to the formula \eqref{DickPol}, the irreducible polynomial over $\F_{2^3}$, for which the generator of the optimal normal basis of II type is $\alpha=t$ is $f(t)=t^{3}+t^2+1$. However, using basic algebraic properties of normal and polynomial bases, we can check that it is true.

So now we will check if for the field $\F_{2^{3}}$ generated by the polynomial $f(t)=t^{3}+t^2+1$, the generator of the normal basis will be the element $\alpha =t$. To do this, note that:
\\\\

\begin{tabular}{|l|l|}
\hline
Normal basis elements & Polynomial basis elements \\ \hline
$\alpha $ & $t$ \\ \hline
$\alpha ^{2}$ & $t^{2}$ \\ \hline
$\alpha ^{4}$ & $t^{4}=t^{2}+t+1$ \\ \hline
\end{tabular}
\\\\

To check if $\alpha$ is indeed the normal basis generator, we will try to create the transition matrix between the normal basis and the polynomial basis.

Assume that $b_{2},b_{1},b_{0}$ are the coefficients of an element in the normal basis (the element is of the form $b_{2}\alpha^{4}+b_{1}\alpha^{2}+b_{0}\alpha$), while $a_{2},a_{1},a_{0}$ are the coefficients in the polynomial basis (the element is of the form $a_{2}t^{2}+a_{1}t+a_{0}$).
Then the transition from the normal basis to the polynomial basis can be done as follows:
\begin{equation*}
    \begin{array}{c}
    \begin{bmatrix}
        a_2 & a_1 & a_0 
    \end{bmatrix}
    \begin{bmatrix}
        t^2 \\
        t \\
        1 \\
    \end{bmatrix}
    =
    \begin{bmatrix}
        b_2 & b_1 & b_0
    \end{bmatrix}
    \begin{bmatrix}
        \alpha^4 \\
        \alpha^2 \\
        \alpha \\
    \end{bmatrix}
    = \\
    \begin{bmatrix}
        b_2 & b_1 & b_0
    \end{bmatrix}
    \begin{bmatrix}
        t^2 + t + 1 \\
        t^2 \\
        t \\
    \end{bmatrix}
    = 
    \begin{bmatrix}
        b_2 & b_1 & b_0
    \end{bmatrix}
    \begin{bmatrix}
        1 & 1 & 1 \\
        1 & 0 & 0 \\
        0 & 1 & 0 \\
    \end{bmatrix}
    \begin{bmatrix}
        t^2 \\
        t \\
        1 \\
    \end{bmatrix}.
    \end{array}
\end{equation*}


From above, it results that:
\begin{equation*}
    \begin{bmatrix}
        a_2 & a_1 & a_0
    \end{bmatrix}
    \begin{bmatrix}
        t^2 \\ t \\ 1
    \end{bmatrix}
    =
    \begin{bmatrix}
        b_2 + b_1 & b_2 + b_0 & b_2
    \end{bmatrix}
    \begin{bmatrix}
        t^2 \\ t \\ 1
    \end{bmatrix} .
\end{equation*}


Thus, it can be immediately noted that $\alpha =t$ will be the generator of the normal basis because the transition matrix from the normal basis to the polynomial basis 
\begin{equation*}
    M_{N\rightarrow P}=
    \begin{bmatrix}
        1 & 1 & 1 \\
        1 & 0 & 0 \\
        0 & 1 & 0 \\
    \end{bmatrix}
\end{equation*}
is non-singular.

The transition matrix from the polynomial basis to the normal basis is:
\begin{equation*}
    M_{P\rightarrow N}=M^{-1}_{N\rightarrow P}=
    \begin{bmatrix}
         0 & 1 & 0 \\
         0 & 0 & 1 \\
         1 & 1 & 1 \\
    \end{bmatrix}.
\end{equation*}

The result of multiplying two elements 
\begin{equation*}
    A=a_2 \alpha^4+a_1 \alpha^2+a_0 \alpha
\end{equation*} and 
\begin{equation*}
    B=b_2 \alpha^4+b_1 \alpha^2+b_0 \alpha
\end{equation*} is:
\begin{equation*}
    \begin{array}{c}
         a_2 b_2 \alpha^8+(a_1 b_2+a_2 b_1)\alpha^6+(a_0 b_2+a_2 b_0)\alpha^5+a_1 b_1 \alpha^4 + \\
         (a_0 b_1+a_1 b_0)\alpha^3+a_0 b_0 \alpha^2.
    \end{array}
\end{equation*}
Note that:
\begin{equation}
    \begin{array}{rl}
    \alpha^8=&t,\\
    \alpha^6=&t^2+t,\\
    \alpha^5=&t+1,\\
    \alpha^4=&t^2+t+1,\\
    \alpha^3=&t^2+1,\\
    \alpha^2=&t^2.
    \end{array}
\end{equation}

Thus, the result of the multiplication can be written using the polynomial basis as
\[
\left(a_1 b_2+(a_2+a_1+a_0) b_1+(a_1+a_0) b_0\right)t^2+
\]
\[
+\left((a_2+a_1+a_0)b_2+(a_2+a_1)b_1+a_2 b_0\right)t+
\]
\[
+a_0 b_2+(a_1+a_0)b_1+(a_2+a_1)b_0=
\]
\[
c'_2 t^2+c'_1 t+c'_0.
\]

Using the transition matrix from the polynomial basis to the normal basis, we obtain that
\[
c'_2 t^2+c'_1 t+c'_0=c'_0 \alpha^4 +(c'_2+c'_0)\alpha^2+c'_1+c'_0=c_2 \alpha^4+c_1\alpha^2+c_0.
\]
Thus, $c_2=c'_0$, $c_1=c'_2+c'_0$, and $c_0=c'_1+c'_0$.
We then get that
\[
c_0=a_2 b_2+a_1 b_2+a_2 b_1+a_0 b_1+a_1 b_0,
\]
\[
c_1=a_0 b_0+a_2 b_0+a_0 b_2+a_1 b_2+a_2 b_1,
\]
\[
c_2=a_1 b_1+a_0 b_1+a_1 b_0+a_2 b_0+a_0 b_2.
\]

Thus, we have obtained the multiplication matrices.
\begin{center}
   $T^{(0)}=\left[
\begin{tabular}{lll}
$0$ & $1$ & $0$ \\ 
$1$ & $0$ & $1$ \\ 
$0$ & $1$ & $1$%
\end{tabular}%
\right]$,  
\end{center}

\begin{center}
 $T^{(1)}=\left[ 
\begin{tabular}{lll}
$1$ & $0$ & $1$ \\ 
$0$ & $0$ & $1$ \\ 
$1$ & $1$ & $0$%
\end{tabular}%
\right]$,    
\end{center}

\begin{center}
   $T^{(2)}=\left[ 
\begin{tabular}{lll}
$0$ & $1$ & $0$ \\ 
$1$ & $0$ & $1$ \\ 
$0$ & $1$ & $1$%
\end{tabular}%
\right]$. 
\end{center}

So now, let us consider the multiplicative subgroup of field $\F_{2^3}$ generated by irreducible polynomial $f(t)=t^3+t^2+1$. Let $g=t$ be the generator of this subgroup. Let $h=t^4+t^2$. We will show how to transform this problem to the QUBO form.

\subsection{Transformation of the example problem}
First, let us look that the order of multiplicative subgroup $\F_{2^3}^{*}$ is equal to $7$, which is prime. Therefore, we have to solve the following problem:
\begin{equation}
    g^y \equiv h \mod(f(t)),
\end{equation}
which is equivalent to the problem of solving
\begin{equation}
    t^y \equiv t^4+t^2 \mod(t^3+t^2+1),
\end{equation}
where $y=4u_2 + 2u_1 + u_0$, for binary variables $u_0, u_1, u_2$. We use normal bases instead of commonly used polynomial bases.

Let us note that it is equivalent to
\begin{equation}
    \begin{array}{c}
        t^{4u_2 + 2u_1 + u_0}=t^{4u_2} t^{2u_1} t^{u_0} = \\ {t^4}^{u_2}{t^2}^{u_1}{t}^{u_0} \equiv t^4+t^2\mod(t^3+t^2+1).
    \end{array} 
\end{equation}

As we use normal basis system representation, one can use vector notation in which $t=[0,0,1]$ and $t^2$ and $t^4$ are simply rotations of $t$. In such a case $t^2=[0,1,0]$ and $t^4=[1,0,0]$. Let us note that the neutral element in normal basis is $t^4+t^2+t=[1,1,1]$.

Let us take the multiplication matrix $T^{(0)}$, defined as:

\begin{equation}
   T^{(0)}= \begin{bmatrix}
$0$ & $1$ & $0$ \\ 
$1$ & $0$ & $1$ \\ 
$0$ & $1$ & $1$
\end{bmatrix}.
\end{equation}

Now let $A,B,C \in \F_{2^3}$, where $C=A \cdot B$. Then, product $C$ is given by

\begin{equation}
\label{EqF23}
\begin{array}{rl}
c_0&=a_2 b_2+a_1 b_2+a_2 b_1+a_0 b_1+a_1 b_0,\\
c_1&=a_0 b_0+a_2 b_0+a_0 b_2+a_1 b_2+a_2 b_1,\\
c_2&=a_1 b_1+a_0 b_1+a_1 b_0+a_2 b_0+a_0 b_2.
\end{array}
\end{equation}

Now, let us note that
\begin{equation}
t^{u_0}=\begin{cases}
            1=t+t^2+t^4, u_0=0,\\
            t, u_0=1.
        \end{cases}    
\end{equation}

So $t^{u_0}$ may be presented as:
\begin{equation}
    t^{u_0}=t+(1-u_0)t^2+(1-u_0)t^4=[1-u_0, 1-u_0, 1].
\end{equation}

Writing $t^{u_0}$ in general form and using new variables, one obtains $t^{u_0}=v_{0,0}t + v_{0,1} t^2 + v_{0,2} t^4$, where $v_{0,0}=1, v_{0,1}=1-u_0, v_{0,2}=1-u_0$.

Now let's perform the multiplication of $t^{u_0}$ by $t^{2{u_1}}$. Similarly as before, if $u_1=0$, then the result will be equal to $t^{u_0}=[v_{0,2}, v_{0,1}, v_{0,0}]$. If $u_1=1$, one must use Equation \eqref{EqF23}.
In such a case the resulting vector $[v_{1,2}, v_{1,1}, v_{1,0}]$
will be of the following form:
\begin{equation}
    \begin{cases}
        v_{1,0}=v_{0,2} u_1+v_{0,0} u_1+(1-u_1)v_{0,0},\\
        v_{1,1}=v_{0,2} u_1+(1-u_1) v_{0,1},\\
        v_{1,2}=v_{0,1}u_1 + v_{0,0} u_1+ (1-u_1) v_{0,2}.
    \end{cases}
\end{equation}

Now, let's note that the system of equations above must be considered while analyzing the transformation of the DLP over $\F_{2^3}$ to the QUBO problem.

Finally, the last step is multiplication of $[v_{1,2}, v_{1,1}, v_{1,0}]$ by $t^{4u_2}$. This step goes as follows:

\begin{equation}
    \begin{cases}
        v_{2,0}=v_{1,2} u_2+v_{1,1} u_2+(1-u_2)v_{1,0}=0,\\
        v_{2,1}=v_{1,0} u_2+v_{1,1} u_2 +(1-u_2) v_{1,1}=1,\\
        v_{2,2}=v_{1,0}u_2 + (1-u_2) v_{1,2}=1.
    \end{cases}
\end{equation}

But let's note that because $h=t^4+t^2$, then the vector $[v_{2,2}, v_{2,1}, v_{2,0}]$ is equal to $[1,1,0]$.

Now, we transform and simplify the system of equations given above.
We set $u_3=v_{0,0}, u_4=v_{0,1}, u_5=v_{0,2}, u_6=v_{1,0}, u_7=v_{1,1}, u_8=v_{1,2}, v_{2,0}=0, v_{2,1}=1, v_{2,2}=1$.

Let us note that first:

\begin{equation}
    \begin{cases}
        u_3=1,\\
        u_4=1-u_0,\\
        u_5=1-u_0=u_4.
    \end{cases}
\end{equation}
Then
\begin{equation}
    \begin{cases}
        u_6=u_5 u_1+u_3 u_1 + (1-u_1) 1=\\
        u_4 u_1+u_3 u_1+1-u_1=u_4 u_1+1,\\
        \\
        u_7=u_4 u_1+(1-u_1)u_4=u_4,\\
        \\
        u_8=u_4 u_1+u_1+(1-u_1)u_4=u_1+u_4.
    \end{cases}
\end{equation}

And then
\begin{equation}
    \begin{cases}
        0=u_2 u_8+u_2 u_7+(1-u_2)u_6=u_2u_8+u_2u_4+(1-u_2)u_6,\\
        1=u_2 u_6+u_2 u_7+(1-u_2)u_7=u_2 u_6 + u_7=u_2 u_6+u_4,\\
        1=u_2u_6+(1-u_2)u_8=u_2 u_6+u_8-u_2u_8.\\
    \end{cases}
\end{equation}

Taking into account the whole equations, making at first linearization, removing all terms on the left side and making squaring, and, finally, adding at the end penalty, one can obtain the final QUBO form of our problem of solving DLP over binary fields. This method is presented in detail, for example, in \cite{Wro22DLP} and \cite{wronski2024transformation}. The necessary equations to prepare the final QUBO problems are given below. Let us note that in the system below, using arithmetic tricks, the number of additional variables necessary to represent the multiplicities of $2$ after transforming the equation from the boolean function to the pseudo boolean function is lowered:
\begin{equation}
    \begin{cases}
F_1=(1-{u_{0}}-{u_{4}})^2,\\
F_2=(1-{u_{6}}-{u_{10}})^2,\\
F_3=({u_{8}}+{u_{1}}+{u_{4}}-2{u_{14}})^2,\\
F_4=(-{u_{11}}+{u_{12}}+{u_{6}}-{u_{13}})^2,\\
F_5=(1-{u_{13}}-{u_{4}})^2,\\
F_6=(1-{u_{13}}-{u_{8}}+{u_{11}})^2,\\
Pen_1=Penalty({u_{1}}, {u_{4}}, {u_{10}}),\\
Pen_2=Penalty({u_{2}}, {u_{8}}, {u_{11}}),\\
Pen_3=Penalty({u_{2}}, {u_{4}}, {u_{12}}),\\
Pen_4=Penalty({u_{2}}, {u_{6}}, {u_{13}}),
    \end{cases}
\end{equation}
where $Pen$ is a standard penalty in Rosenberg form: when is made substitution $z=xy$, the resulting penalty is of the form $(xy-2(x+y)z+3z)$ and is obtained by invoking function $Penalty(x,y,z)$. Then, the final QUBO problem is given by 
{\small\begin{equation}
\label{QUBO_prob}
F=F_1+F_2+F_3+F_4+F_5+F_6+Pen_1+Pen_2+Pen_3+Pen_4.
\end{equation}}


We transformed the problem above into a QUBO problem using only $11$ logical variables. Using quantum annealing, we obtained the correct solution of $y=5$.

\begin{figure}[h]
        \centering
        \includegraphics[scale=0.75]{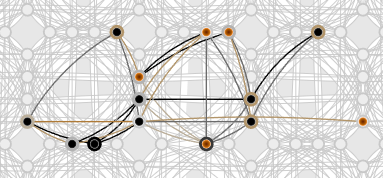}
        \caption{Embedding of DLP for $\F_{2^3}$ in the D-Wave Advantage 2 QPU.}
        \label{fig1}
    \end{figure}

The scheme presenting embedding of our problem into the D-Wave Advantage 2 prototype 2.3 system having 1248 qubits is presented in Figure \ref{fig1}.

The problem and used solver parameters are presented in Table \ref{tab1}.

\begin{table}[H]
\begin{center}
\begin{tabular}{|l|l|}
\hline
Parameter												& Value				\\ \hline
Name (chip ID)                                          & Advantage2$\_$prototype2.3 \\ \hline
Available qubits			                                        & 1,248		\\ \hline
Topology		                                        & Zephyr		\\ \hline
Number of reads                                         & 10,000	\\ \hline
Annealing time ($\mu s$)                                & 20			\\ \hline
Anneal schedule                                         & [[0,0],[20,1]]	\\ \hline
H gain schedule                                         & [[0,0],[20,1]]		\\ \hline
Programming thermalization ($\mu s$)                    & 1000  			\\ \hline
\end{tabular}
\caption{D-Wave Advantage solver parameters used in solving QUBO problem equivalent to the problem of finding discrete logarithm over $\mathbb F_{2^3}$ in the subgroup of size $7$.}
\label{tab1}
\end{center}
\end{table}

\begin{table}[H]
\begin{center}
\begin{tabular}{|l|l|}
\hline
Parameter												& Value	\\ \hline
 Number of target variables                              & 14	\\ \hline
Max chain length                                        & 2				\\ \hline
 Chain strength                                          & 1.8827			\\ \hline
QPU access time ($\mu s$)                               & 804,627.61				\\ \hline
QPU programming time ($\mu s$) 							& 19,227.61		\\ \hline
 QPU sampling time ($\mu s$) 							& 785,400		\\ \hline
 Total post-processing time ($\mu s$) 					& 1	\\ \hline
Post processing overhead time ($\mu s$)                 & 1		\\ \hline
\end{tabular}
\caption{Results parameters obtained during solving QUBO problem equivalent to the problem of finding discrete logarithm over $\mathbb F_{2^3}$ in the subgroup of size $7$.}
\label{tab2}
\end{center}
\end{table}

We ran the problem above using quantum annealing 10,000 times. 7,415 trials gave proper minimal energy, which means that for the given example, the probability of obtaining the proper result is equal to 74.15\%.

\section{Probability analysis}

In this section we will analyze the probability of obtaining correct solution using quantum annealing. We will show that, contrary to Shor's algorithm application, in our case it is much easier to show that our results are not random.

It is worth to note that experiments of solving small DLPs instances over prime fields have been conducted using Shor's algorithm. As far as we know, the biggest succesfully solved experiment \cite{Yos22} was solving:

\begin{equation}
    2^x \equiv 1 (mod\ 3).
\end{equation}

However, even for such small example it was not trivial to show that application of Shor's algorithm worked correctly and wasn't the result of random computations.

We begin by calculating how many minimal solutions our QUBO problem \eqref{QUBO_prob} has.

As in mamy cases there are possible many equivalent proper solutions, in our case there exists only one proper solution. Let us note that because the proper $y=5$, it means that it is the only proper solution of exponent. It is worth to note that if proper solution would be $y=0$, then also $y=7$ would be proper solution of our problem, because $7 \mod{7}=0$.

Now we have to check if there may be equivalent proper solutions, where only values of multiplicities $k's$ differ. It is worth to note that if for example the some value $k_i$ may be at most equal to $2$, then in such a case $k=u_i + u_j$ for some indices $i,j$ and binary variables $u_i$ and $u_j$. But let's note that if proper solution is $k=1$, then such solution may be obtained on two different ways. The first solution will be for $u_i=1, u_j=0$, while the second will be for $u_i=0, u_j=1$.

However, let us look that in the definition of our QUBO problem there is only one multiplicity $k$ (in our case it is only $u_{14})$, and it is written using only $1$ bit. Therefore, only one minimal energy solution is in our case possible.

Now we will show that running our experiment using quatnum annealing, obtaining the minimal energy solution randomly is practically impossilble. So we will estimate the probability of obtaining proper solution using quantum annealing. Let's note that, using for example Shor's algorithm, the space is much smaller and therefore, the probability distribution of different possible values (proper and also unproper) looks more uniformly. Therefore, in Shor's algorithm often a lot of effort goes to analysis if obtaining proper solution has appropriate big probability and if it may be distinguished from the random solution.

Below we prove that our solution is not obtained randomly.

Let's note that our QUBO problem consists of 11 logical variables. It means that the solution space is $2^{11}=2048$. In this case, only one solution is proper, which means that, using binomial distribution, the probability of single success is equal to $\frac{1}{2048}$. 

We made $10,000$ trials. Assuming that from the cryptographic point of wiev we may assume that method works if returns the proper solution with probability $0.5$, we will compute how much likely is that one obtains at least 5000 successes, making 10000 trials, if obtaining each state would be as same probable. In such a case one has to compute cumulated binomial distribution
\begin{equation}
    \sum_{i=5000}^{10000}{{{1000}\choose{i}} \left(\frac{1}{2048}\right)^{i} \left(\frac{2047}{2048}\right)^{10000-i}}.
\end{equation}

It is worth noting that such probability is extremely small and is equal to approximately $3.103 \cdot 10^{-13550}$.

As in our experiment we obtained 7,415 successes in 10,000 trials, it is clear that our experiment is not a random one, therefore it is statistically proved that our method works correctly and gives almost $75\%$ probability of success.

\section{Conclusion}

In this paper the application of quantum annealing to solve discrete logarithm problem over $\F_{2^n}$ fields has been presented. It is worth to note that our method of transformation od DLP to the QUBO problem requires approximately $3n^2$ logical variables if for given field the optimal normal basis exists.
We presented the experiment where DLP over small binary field ($\F_{2^3}$) was solved using quantum annealing. Our experiment has almost $75\%$ probability of returning proper result for such small field.

However existing of optimal normal basis in the given field is restrictive assumption, one should note that for many binary fields such optimal normal bases exist. If for a given field such optimal normal basis does not exist, then the presented method of transformation of the DLP to the QUBO problem would require asymptotically more resources.

It is however unknown how far we can go on the current quantum annealers, because we did not conducted such experiments for larger fields. Further works should cover this gap and solving of larger instances be tried.

\bibliographystyle{splncs04}
\bibliography{bibliography}

\end{document}